\documentclass[aps,prd,twocolumn,showpacs,floats,floatfix,letterpaper,nofootinbib,superscriptaddress,]{revtex4}

\usepackage{amssymb,amsmath,latexsym,mathrsfs}
\usepackage{graphicx}
\usepackage{epsfig}
\usepackage{multirow}
\usepackage{array}
\usepackage{color}

\newcommand{\mnu}{{\Sigma}m_{\nu}}

\begin{document}


\title{Constraints on neutrino masses from Planck and Galaxy Clustering data}
\author{Elena Giusarma}
\affiliation{IFIC, Universidad de Valencia-CSIC, 46071, Valencia, Spain}
\author{Roland de Putter}
\affiliation{Jet Propulsion Laboratory, California Institute of Technology, Pasadena, CA 91109
\& California Institute of Technology, Pasadena, CA 91125}
\author{Shirley Ho}
\affiliation{McWilliams Center for Cosmology, Department of Physics,
Carnegie Mellon University, 5000 Forbes Ave., Pittsburgh, PA
15213}
\author{Olga Mena}
\affiliation{IFIC, Universidad de Valencia-CSIC, 46071, Valencia, Spain}
\begin{abstract}
We present here bounds on neutrino masses from the combination of recent
Planck Cosmic Microwave Background measurements and galaxy clustering
information from the Baryon Oscillation Spectroscopic Survey (BOSS), part
of the Sloan Digital Sky Survey-III. We use the full shape of either the
photometric angular clustering (Data Release 8) or the 3D spectroscopic
clustering (Data Release 9) power spectrum in different cosmological
scenarios. In the $\Lambda$CDM scenario, spectroscopic galaxy clustering measurements 
 improve significantly the existing neutrino mass bounds from Planck
data. We find $\sum m_\nu< 0.39$~eV at $95\%$ confidence level for the combination of the 3D
power spectrum with Planck CMB data (with lensing included) and
Wilkinson Microwave Anisoptropy Probe 9-year polarization
measurements. Therefore, robust neutrino mass constraints can be obtained
without the addition of the prior on the Hubble constant from HST.

In extended cosmological scenarios with a dark energy
fluid or with non flat geometries, galaxy clustering measurements are
essential  to pin down the neutrino mass bounds, providing in the majority of cases better results than those obtained
from the associated measurement of the Baryon Acoustic Oscillation scale
only. In the presence of a freely varying (constant) dark energy equation of
state, we find $\sum m_\nu<0.49$~eV at $95\%$ confidence level for the combination of the 3D
power spectrum with Planck CMB data (with lensing included) and
Wilkinson Microwave Anisoptropy Probe 9-year polarization measurements. This same data combination 
in non flat geometries provides the neutrino mass bound $\sum m_\nu<0.35$~eV at $95\%$ confidence level.

\end{abstract}


\maketitle

\section{Introduction}
Massive neutrinos leave distinct imprints in the different cosmological
data sets. Concerning Cosmic Microwave Background (CMB) anisotropies, 
the primary effect of neutrino masses is via the \emph{Early Integrated Sachs
Wolfe effect}. The transition from the relativistic to the non
relativistic neutrino regime will affect the decays of the
gravitational potentials at the decoupling period, leading to a non
negligible signature around the first peak. This has been,
traditionally, the most relevant signature from neutrino masses on the
CMB~\cite{sergio}. However, recent neutrino mass bounds from Planck data~\cite{planck}, seem to be driven by the 
massive neutrino signature on gravitational lensing. A non zero value of the
neutrino mass will induce a higher expansion rate which  will suppress
the clustering on scales smaller than the horizon while neutrinos turn
non relativistic~\cite{lensingnu}.  Regarding large scale structure, due to the large neutrino velocity
dispersion, the non-relativistic neutrino overdensities will only cluster at wavelengths larger than their
free streaming scale. Consequently, the growth of matter density
perturbations is reduced and the matter power spectrum is suppressed
at small scales, see Ref. \cite{sergio2} and references therein. Therefore, cosmological data provide a unique tool to test the
neutrino masses, see
Refs.~\cite{Reid:2009nq,Hamann:2010pw,dePutter:2012sh,Giusarma:2012ph,Zhao:2012xw,Hinshaw:2012fq,Hou:2012xq,
Sievers:2013wk,Archidiacono:2013lva} for neutrino mass bounds before Planck CMB data release.

The limits from Planck satellite, including lensing as well as low-$\ell$
polarization measurements from WMAP 9-year data~\cite{Bennett:2012fp}
(WP) are $\sum m_\nu<1.11$~eV at $95\%$~CL. The addition of a prior on the Hubble constant $H_0$ from
the Hubble Space Telescope~\cite{Riess:2011yx} improves the constraint in a very
significant way, $\sum m_\nu<0.21$~eV. This is due to the strong
degeneracy between $H_0$ and $\sum m_\nu$ at $95\%$~CL: if the sum of the neutrino
masses is increased, the change induced in the distance to last
scattering can be compensated by lowering $H_0$~\cite{Giusarma:2012ph}.
However, Planck and HST measurements of the Hubble constant $H_0$
show a $2.5 \sigma$ tension and therefore, it is fortunate that datasets
other than the HST prior may help in pinning down the bound on neutrino mass from CMB data alone. 

Baryon Acoustic Oscillation (BAO) data, as measured by the
Sloan Digital Sky Survey (SDSS) Data Release 7~\cite{dr71,dr72},
the WiggleZ survey~\cite{wigglez}, the Baryon Acoustic Spectroscopic Survey
(BOSS) SDSS-III Data Release 9~\cite{anderson} and 6dF~\cite{6df},
also significantly improve the
constraints, leading to $\sum m_\nu<0.26$~eV at $95\%$~CL when combined with Planck (with
lensing) and WP data. However, in non minimal scenarios with a curvature or with a
more general dark energy component these constraints are notably
degraded and \emph{geometrical} BAO information from galaxy clustering may
not be as powerful as \emph{shape} measurements of the matter power
spectrum. Previous works~\cite{Hamann:2010pw,Giusarma:2012ph} have noticed the advantages of using
full power spectrum measurements in extended cosmological scenarios
due to their ability of removing degeneracies. 

Here we combine recent Planck data with galaxy power spectrum measurements
from the BOSS experiment~\cite{Dawson:2012va}, one of the four surveys of
the Sloan Digital Sky Survey III,
SDSS-III~\cite{Eisenstein:2011sa}. 
We consider first the 2D angular power spectrum measurements~\cite{Ho:2012vy} from the CMASS sample~\cite{White:2010ed} of luminous
galaxies of SDSS Data Release 8 (DR8)~\cite{Aihara:2011sj}. We then explore
as well the neutrino mass constraints from the full 3D power spectrum
shape of SDSS Data Release 9 (DR9) ~\cite{Ahn:2012fh}. While DR8
contains the full photometric CMASS sample, DR9 provides the galaxy
spectra of CMASS galaxies, the largest publicly available set of galaxy spectra to date.

The authors of  Ref.~\cite{dePutter:2012sh}, in the context of a $\Lambda$CDM model,
found $\sum m_\nu<0.36$~eV ($\sum m_\nu<0.26$~eV) at $95\%$~CL 
with (without) shot noise-like parameters when combining WMAP 7 year data
with DR8 2D angular power spectrum measurements plus a HST prior on $H_0$. 
Exploiting DR9 3D power spectrum measurements Ref.~\cite{Zhao:2012xw} quotes the bound $\sum
m_\nu<0.34$~eV at $95\%$~CL after combining with WMAP7, supernova data
and additional BAO measurements within a $\Lambda$CDM model. 

We shall update here the constraints quoted above, quantifying the benefits
from the improved CMB Planck data. Our neutrino mass constraints are
presented in  different fiducial cosmologies, namely, non flat and
dynamical dark energy cosmologies. We also show the impact on our constraints of 
the underlying galaxy power spectrum, adopting different models
to describe galaxy clustering. 

The structure of the paper is as follows. In Sec.~\ref{sec:params} we describe
the parameters used in the analysis. Planck CMB and galaxy clustering
data, plus galaxy clustering modeling are described in
Sec.~\ref{sec:data}. Section~\ref{sec:results} contains our results, and
we draw our conclusions in Sec.~\ref{sec:concl}.

\section{Cosmological parameters}
\label{sec:params}
The standard, three massive neutrino scenario we explore here is
described by the following set of parameters:
\begin{equation}
\label{parameter}
  \{\omega_b,\omega_c, \Theta_s, \tau, n_s, \log[10^{10}A_{s}], \sum m_\nu\}~,
\end{equation}
$\omega_b\equiv\Omega_bh^{2}$ and $\omega_c\equiv\Omega_ch^{2}$  
being the physical baryon and cold dark matter energy densities,
$\Theta_{s}$ the ratio between the sound horizon and the angular
diameter distance at decoupling, $\tau$ is the reionization optical depth,
$n_s$ the scalar spectral index, $A_{s}$ the amplitude of the
primordial spectrum and $\sum m_\nu$ the sum of the masses of the
three active neutrinos in eV. We assume a degenerate neutrino mass
spectrum in the following. The former scenario is enlarged with $w$
and $\Omega_k$ in the case of extended models.
Table \ref{tab:priors} specifies the priors considered on the different cosmological
parameters.
For our numerical analyses, we have used the Boltzmann CAMB
code~\cite{camb} and 
extracted cosmological parameters from current data
using a Monte Carlo Markov Chain (MCMC) analysis based on the publicly available MCMC package \texttt{cosmomc}~\cite{Lewis:2002ah}.

\begin{table}[h!]
\begin{center}
\begin{tabular}{c|c}
\hline\hline
 Parameter & Prior\\
\hline
$\Omega_{b}h^2$ & $0.005 \to 0.1$\\
$\Omega_{c}h^2$ & $0.01 \to 0.99$\\
$\Theta_s$ & $0.5 \to 10$\\
$\tau$ & $0.01 \to 0.8$\\
$n_{s}$ & $0.9 \to 1.1$\\
$\ln{(10^{10} A_{s})}$ & $2.7 \to 4$\\
$\sum m_\nu$ (eV) &  $0.06 \to 3.$\\
$\Omega_k$ &  $-0.3 \to 0.3$\\
$w$ &  $-2 \to 0$\\

\hline\hline
\end{tabular}
\caption{Uniform priors for the cosmological parameters considered here.}
\label{tab:priors}
\end{center}
\end{table}

\section{CMB and Galaxy Clustering measurements}
\label{sec:data}
\subsection{Planck}
We consider the high-$\ell$ TT likelihood, including measurements up to a
maximum multipole number of $\ell_{\rm max}=2500$, combined with the
low-$\ell$ TT likelihood, including measurements up to $\ell=49$ and the 
low-$\ell$ WMAP TE,EE,BB likelihood including multipoles up to $\ell=23$. 
We include the lensing likelihood in all our Monte Carlo analyses.
We refer to this data set as the PLANCK data set.

We also consider the effect of a gaussian prior on the Hubble constant
$H_0=73.8\pm2.4$ km/s/Mpc, accordingly with the measurements from the
Hubble Space Telescope~\cite{Riess:2011yx}. We refer to this prior as HST.

We fix the helium abundance to $Y_p=0.24$ and the lensing spectrum
normalization to $A_L=1$.  We marginalize over all foregrounds parameters as described in \cite{planck}.

\subsection{DR8 Angular Power Spectrum}
\subsubsection{DR8 Data}

We exploit the stellar mass-limited DR8 CMASS sample of luminous 
galaxies, detailed in \cite{White:2010ed},  divided into four 
photometric redshift bins, $z=0.45-0.5-0.55-0.6-0.65$. 
The photometric redshift error lies within the range $\sigma_z(z) = 0.04 - 0.06$,
increasing from low to high redshift, see Refs. \cite{rossetal11,Ho:2012vy}.
The calculation of the angular power spectrum for each bin 
is described in detail in Ref.~\cite{Ho:2012vy}. The expectation value
of the power spectrum is a convolution of the true power spectrum with a window function, see
\cite{seobao} for examples on these window functions.
When fitting the data to the underlying theoretical model, 
we always apply these window functions to the theoretical power
spectra before calculating the likelihood relative to the data. 
To avoid large systematic uncertainties~\cite{rossetal11,Ho:2012vy} we do not consider bands with $\ell
< 30$ in our analysis. 

\begin{table}[h!]
\begin{center}
\begin{tabular}{c|c}
\hline\hline
 DR8 parameters & Prior\\
\hline
$b_i$ & $0.5 \to 5$\\
$a_i$& $-5 \to 12$\\

\hline\hline
\end{tabular}
\caption{Uniform priors for the DR8 bias and shot noise parameters in
each of the four photometric redshift bins $z=0.45-0.5-0.55-0.6-0.65$ used in the
DR8 clustering data analyses.}
\label{tab:dr8}
\end{center}
\end{table}
\subsubsection{DR8 Clustering model}

In order to describe the theoretical angular power spectrum, we follow
here the simple linear scale independent bias model described in
Ref.~\cite{dePutter:2012sh}, characterized by four free bias
parameters $b_i$ (i.e. one per 
each redshift bin). In addition to these bias parameters,
we also consider shot noise-like parameters $a_i$
\begin{eqnarray}
\label{eq:clfull2}
C_\ell^{(ii)}& =& b_i^2 \, \frac{2}{\pi} \int k^2 dk \, P^{\rm
  m}(k,z=0) \, \times \nonumber \\
&&\left( \Delta_\ell^{(i)}(k) + \Delta^{{\rm RSD}, (i)}_\ell(k)
\right)^2 + a_i~,
\end{eqnarray}
where the $a_i$ parameters mimic the effects of a scale-dependent
galaxy bias as well as the effect of potential insufficient shot noise
subtraction. 
Table~\ref{tab:dr8} denotes the priors adopted on the bias and
shot noise parameters in each of the four redshift bins exploited
here. The neutrino mass bounds presented in the next section  will
be derived by default including the shot noise parameters
$a_i$ in the next section, although we shall mention on some cases the
bounds without shot noise. In Eq.~(\ref{eq:clfull2}), $P^{\rm m}(k,z=0)$ is the matter power spectrum at redshift zero after
applying the HaloFit prescription~\cite{Smithetal03,Takahashi:2012em} to account for
non-linear effects~\footnote{Although the revisited version of the HaloFit
  model~\cite{Takahashi:2012em} accounts for a constant, $w\neq-1$
  dark energy equation of state, it is restricted to flat
  models. In principle, a linear interpolation of the fitting functions to
  N-body simulations could account for not flat models~\cite{Smith:2002dz}. In practise, since these fitting
  functions have been shown to have a $5-10\%$ discrepancy with
  simulations even in the simplest case of a flat $\Lambda$CDM scheme,
  (see e.g. Ref.~\cite{Vanderveld:2012ec} and references therein),
  we neglect here the extra corrections in the HaloFit
  description in not flat cosmologies.}  and
\begin{eqnarray}
\label{eq:leg}
\Delta_\ell^{(i)}(k) = \int dz \, g_i(z) \, T(k, z) \, j_\ell(k \, d(z))~.
\end{eqnarray}
Here, $g_i(z)$ is the normalized redshift distribution of galaxies in
bin $i$, $j_\ell$ is the spherical Bessel function, $d(z)$
is the comoving distance to redshift $z$ and $T(k, z)$ the
matter transfer function relative to redshift
zero. The contribution due to redshift space distortions is
\begin{eqnarray}
\label{eq:legrsd}
\Delta^{{\rm RSD}, (i)}_l(k) &=& \beta_i \, \int dz \, g_i(z) \, T(k,
z) \, \times \nonumber\\ 
  && \left[\frac{(2l^2+2l-1)}{(2l+3)(2l-1)} j_l(kd(z)) \right. \nonumber \\
   &&\left.- \frac{l(l-1)}{(2l-1)(2l+1)} j_{l-2}(kd(z)) \right. \nonumber \\
   &&\left.- \frac{(l+1)(l+2)}{(2l+1)(2l+3)} j_{l + 2}(k d(z)) \right]~,
\end{eqnarray}
where $\beta_i(z) = f(z)/b_i$ is the redshift distortion parameter and
\begin{equation}
f(z) \equiv \frac{d\ln D(z)}{d\ln a}
\end{equation}
is the growth factor (with $D(z)$ the linear growth function). When
massive neutrinos are an additional ingredient in the universe's mass
energy-density, the growth function is scale-dependent. Following Ref.~\cite{dePutter:2012sh}, we shall ignore the scale 
dependent growth in $\beta(z)$ since it is a small ($\ll 10 \%$) correction to the already small effect of redshift space distortions.

As previously stated, $\ell> 30$ in our data analyses. We consider
$\ell_{\rm max} = 200$, 
value which ensures the suppression of the uncertainties from
non-linear corrections to the modeled angular power
spectra~\cite{dePutter:2012sh}.  For the likelihood function, we use 17 data points per redshift slice. 

 \subsection{DR9 Power Spectrum}
\begin{table}[h!]
\begin{center}
\begin{tabular}{c|c}
\hline\hline
 DR9 parameters & Prior\\
\hline 
$S$&$-1 \to 1$\\
$b_{\rm HF}$  &$0.1 \to 10$\\
$P^{\rm s}_{\rm HF}$ &$100\to 10000$\\
$b_{\rm Q}$ &$0. \to 10$\\
$Q$ &$0.1 \to 100$\\

\hline\hline
\end{tabular}
\caption{Uniform priors for the DR9 bias and shot noise parameters
  $b_{\rm HF}$ and $P^{\rm s}_{\rm HF}$ respectively, in
the case of the HaloFit prescription for the galaxy power spectrum as well as for  $b_{\rm Q}$ and $Q$, free parameters of
the model of Ref.~\cite{Cole:2005sx}. We explore the neutrino mass
constraints for these two galaxy clustering models in the case of the DR9
3D power spectrum.}
\label{tab:priors1}
\end{center}
\end{table}
 \subsubsection{DR9 Data}
Here we use the DR9 CMASS sample of galaxies~\cite{Ahn:2012fh} which
contains $264\,283$ massive galaxies covering 3275\,deg$^2$ with
redshifts $0.43<z<0.7$ (being the effective redshift $z_{\rm eff}=0.57$).
The measured galaxy power spectrum $P_{\rm meas}(k)$ is the one used
in Refs.~\cite{Zhao:2012xw,anderson,beth2,samushia,sanchez,nuza,sys1,sys2}, which is obtained using the standard Fourier
technique~\cite{fkp}, see \cite{beth} for details. This galaxy
power spectrum was the one used to fit the Baryon Acoustic Oscillations~\cite{anderson}. 

On large scales, we are affected by systematic effects from stars or
seeing of the survey. On small scales, we are affected by
observational effects such as redshift failures and fiber collisions. A conservative approach has been provided by
Refs.~\cite{sys1,sys2}, which add an extra free parameter in the measured power spectrum
\begin{equation}
P_{\rm meas}(k) = P_{\rm meas,w}(k)-S[P_{\rm meas,nw}(k)-P_{\rm meas,w}(k)],
\label{eq:measuredpower}
\end{equation}
where $P_{\rm meas,w}(k)$ refers to the measured power spectrum after
applying the weights for stellar density, which represent the main source of
systematic errors, $P_{\rm meas,nw}(k)$ is the measured power
spectrum without these weights and $S$ is an extra nuisance parameter
to be marginalized over, see Tab.~\ref{tab:priors1}. The expectation value of the matter power spectrum
is a convolution of the true matter power spectrum with the window
functions, which account for the correlation of data at different scales
$k$ due the survey geometry. Therefore, the theoretical power
spectra  $P^g_{\textrm{th}}(k)$ (to be computed in the following section)
needs to be convolved with a window matrix before comparing it to 
$P_{\rm meas}(k)$.
In order to avoid non linearities, we adopt the conservative choice of
a  maximum wavenumber of $k_{\rm{max}}=0.12$~$h$/Mpc, region which is
safe against large non linear corrections in the modeled theoretical
spectra, that we discuss below.  We use therefore 22 points in the
range $0.03$~$h$/Mpc $<k<0.12$~$h$/Mpc from the total 74 points of the DR9
power spectrum.

\subsubsection{DR9 Clustering model}
We follow here two different approaches to model the theoretical power
spectrum in the weakly nonlinear regime explored here ($k_{\rm{max}}=0.12$~$h$/Mpc). These
two models are among the three ones considered in
Ref.~\cite{Zhao:2012xw}, where it was checked that the
neutrino mass bounds show a very mild dependence on the galaxy
clustering models considered in their analyses. The first approach we 
consider for DR9 is the HaloFit prescription (HF)~\cite{Smithetal03,Takahashi:2012em}. 
The final theoretical galaxy power spectrum to be convolved with the
window functions reads
\begin{equation}
  P^g_{\textrm{th}}(k, z) = b_{\rm HF}^{2 }P^{\rm m }_{\rm HF\nu}(k; z) +
  P^{\rm s}_{\rm HF}~,
\label{eq:halofit}
\end{equation} 
where $b_{\rm HF}$ and $P^{\rm s}_{\rm HF}$ are the bias and the shot
contribution respectively, considered to be constant. The priors
adopted in the the former two parameters are depicted in Tab. ~\ref{tab:priors1}.
The model given above by Eq.~(\ref{eq:halofit}) with a bias and a shot
noise parameter is equivalent to that used before for modeling the
theoretical angular power spectra ofr DR8 data analyses, see Eq.~(\ref{eq:clfull2}).

The second approach adopted here for galaxy clustering modeling is
that of Ref.~\cite{Cole:2005sx}: 
 \begin{equation}
  P^g_{\textrm{th}}(k, z)= b_{\rm Q}^{2} \frac{1+Qk^2}{1+1.4k} P^{\rm {m, linear}} (k,z)~,
\label{eq:cole}
\end{equation} 
where $k$ is the wavenumber in units of h/Mpc and $P^{\rm {m,
    linear}}$ is the linear matter power spectrum. The free parameters of
this model are  $b_{\rm Q}$ and $Q$, which mimic the scale dependence of the
power spectrum at small scales. These two parameters are considered
here constants with priors specified in Tab. ~\ref{tab:priors1}. 
In the following section we shall comment on the dependence of
the neutrino mass constraints on the underlying galaxy power spectrum
model.

\section{Results} 
\label{sec:results}
Here we present the constraints from current cosmological data sets on
the sum of the three active neutrino masses
$\sum m_\nu$ in different scenarios and with different combinations of
data sets. 
\subsection{Standard Cosmology plus massive neutrinos}
\label{sec:st}
Throughout this section we shall assume a $\Lambda$CDM cosmology, and
compute the bounds on the sum of the three active neutrino masses arising from
the different cosmological data sets considered
here. Table~\ref{tab:constraints1} shows the $95\%$~CL upper bounds on
the total  neutrino mass for PLANCK, PLANCK plus DR8 and PLANCK plus
DR9 data sets, with and without the HST prior on the Hubble constant. 
These limits include the shot noise additional parameters in the
case of DR8 and the systematic effects, in the case of DR9. Notice first that the
constraints from the PLANCK data set described before (which include
the Planck lensing likelihood as well as WMAP 9 year data polarization
measurements) are not very promising, since in this
case $\sum m_\nu < 1.11$~eV at $95\%$~CL. The fact that CMB alone does
not provide very significant constraints on the sum of the neutrino
masses has been already discussed in the literature (see, for
instance~\cite{Giusarma:2012ph}). 
Indeed, without the $H_0$ prior, the change induced in the CMB temperature anisotropies caused by an increase in $\sum m_\nu$ can be
compensated by a decrease in the Hubble constant $H_0$.  An increase in $\sum
m_\nu$ will induce a shift in the distance to last scattering\footnote{$r_{\theta}(z_{\rm rec}) \propto \int_0^{z_{\rm rec}} dz \, \left[\omega_r a^{-4} + \omega_m a^{-3} + (1 - \omega_m/h^2) \right]^{-1/2}$,
with $\omega_m = \omega_b + \omega_c + \omega_\nu$}.
While the acoustic peak structure of the CMB data does not leave much freedom in $\omega_c$ and $\omega_b$,
the change in distance to last scattering could be compensated by
lowering $H_0$. 
The presence of the HST prior on the Hubble parameter will break this strong degeneracy,
setting a $95\%$~CL bound of $0.22$~eV in the sum of the three active neutrino masses.

 However, and as discussed in the introductory section, HST and Planck
 data sets show a tension of $\sim 2.5\sigma$ in their measured value of
 the Hubble constant $H_0$. It is therefore mandatory to explore
 whether other data sets could also strengthen the constraint on $\sum
 m_\nu$ from the PLANCK data set alone. DR8 angular power spectrum measurements, if
 combined with the PLANCK data set, provide an upper limit of $\sum m_\nu <
 0.98$~eV at $95\%$~CL with the shot noise parameters included in the analysis. If we consider instead
 the DR8 BAO angular diameter distance constraint  $D_A(z) = 1411\pm 65$~Mpc at $z =
0.54$~\cite{seobao} and combine this measurement with the PLANCK data set,
 the bound is  $\sum m_\nu < 0.85$~eV at $95\%$~CL. The neutrino
 mass bound from DR8 BAO-only is mildly stronger than the one obtained
 with the full shape of the DR8 galaxy clustering matter spectrum due
 to the larger value of $\ell_{\textrm{max}}=300$ used in the analysis
 of Ref.~\cite{seobao} to extact the angular BAO signature.

When considering the DR9 data set combined with PLANCK, we achieve a bound of $\sum m_\nu< 0.39$~eV at $95\%$~CL. The former limit
is obtained in the case in which the theoretical power spectrum for DR9
is given by Eq.~(\ref{eq:halofit}) which uses the HF prescription. Very
similar bounds are obtained if we use for the theoretical DR9 spectrum
the approach given by Eq.~(\ref{eq:cole}).

If instead of using the
full shape information from BOSS DR9 we
use the DR9 BAO signature~\cite{anderson}, the neutrino mass limit is $\sum m_\nu<0.40$~eV
at $95\%$~CL.  Note that the bound on $\sum m_\nu$ arising from the geometrical BAO DR9 geometrical information
is very similar to that obtained using the full shape of the DR9 3D
clustering measurements. While in the context of the minimal $\Lambda$CDM model, BAO
measurements and galaxy clustering data should provide similar
constraints,  the BAO DR9 signal is extracted
using the matter power spectrum in the range $0.02$~$h$/Mpc
$<k<0.3$~$h$/Mpc~\cite{anderson},  a much wider range than the one
considered in the full power spectrum case. 

To summarize, galaxy clustering data, and, especially, DR9 3D power
spectrum data, helps enormously in improving the neutrino mass
constraints, arriving at $m_\nu<0.39$~eV at $95\%$~CL without the
addition of the measurement of $H_0$ from the HST experiment.
The former bound is not as tight as the value quoted by the Planck
collaboration $\sum m_\nu <0.26$~eV at $95\%$~CL, obtained after combining Planck
measurements (including lensing) with WP and BAO data.  The reason for the
difference among these two $95\%$~CL neutrino mass bounds (i.e. $\sum m_\nu<0.39$~eV versus
$\sum m_\nu < 0.26$~eV) is due to the fact that here we are
considering exclusively BAO information
from DR9 SDSS data, while in 
the Planck analysis other available BAO
measurements have been considered as well.

\begin{table*}
\begin{center}
\begin{tabular}{lccccc}
\hline \hline
      &             Planck+WP+lensing    &       Planck+WP +lensing   &   Planck+WP+lensing\\
      &             (+HST)          &      +DR8 (+HST)  &  +DR9 (+HST)\\
\hline
\hspace{1mm}\\
${\mnu}[eV]$  & $<1.11$  $(0.22)$ & $<0.98$  $(0.23)$ & $<0.39$  $(0.23)$ \\
\hline
\hline
\end{tabular}
\caption{$95\%$~CL upper bounds on ${\mnu}$ in a $\Lambda$CDM model from the different data combinations
  considered here, with (without) the HST prior on the Hubble constant
$H_0$. The results with DR8 (DR9) data sets include the shot
noise (the systematic corrections) parameters.}
\label{tab:constraints1}
\end{center}
\end{table*}

\begin{figure*}
\begin{tabular}{c c}
\includegraphics[width=9cm]{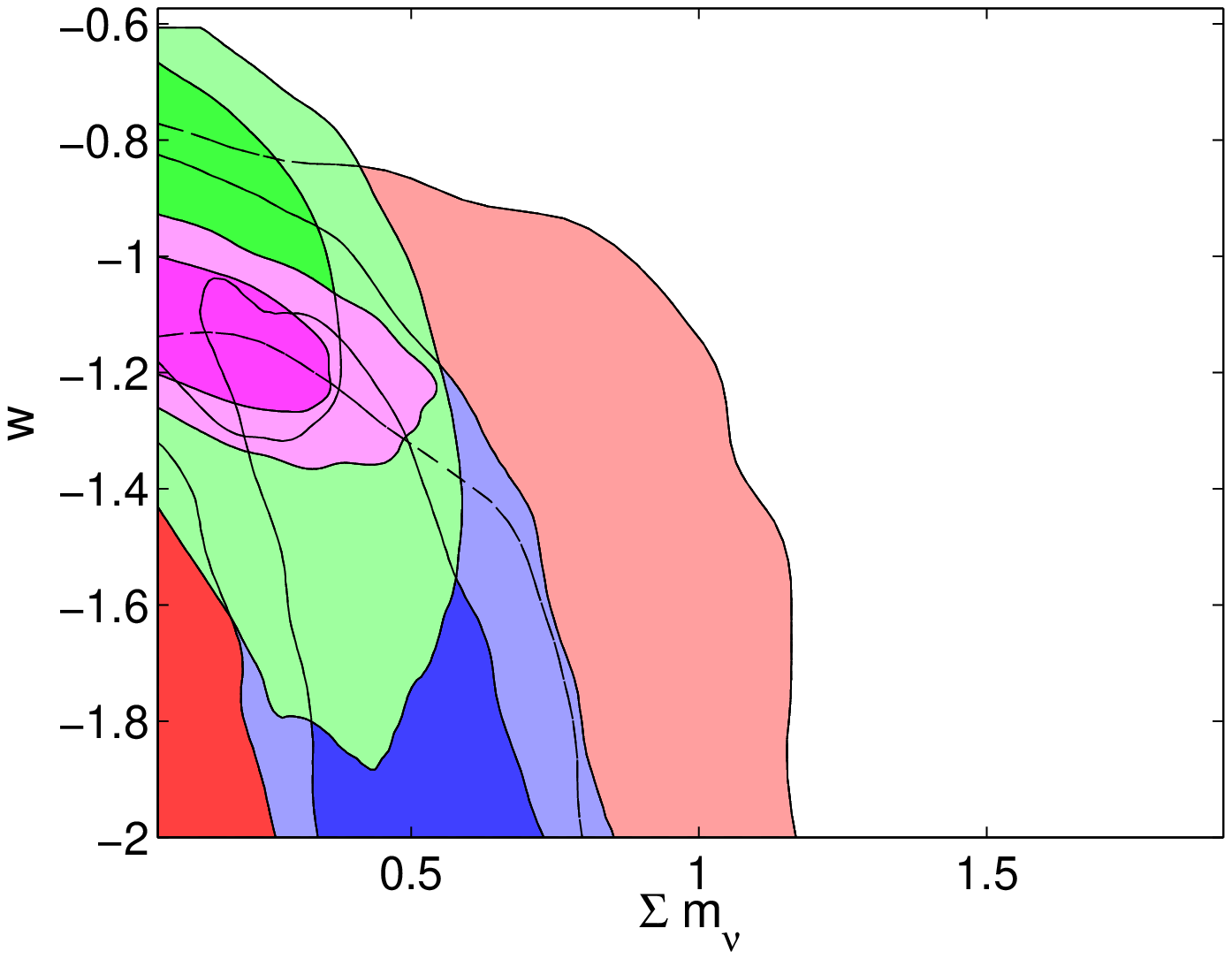}&\includegraphics[width=9cm]{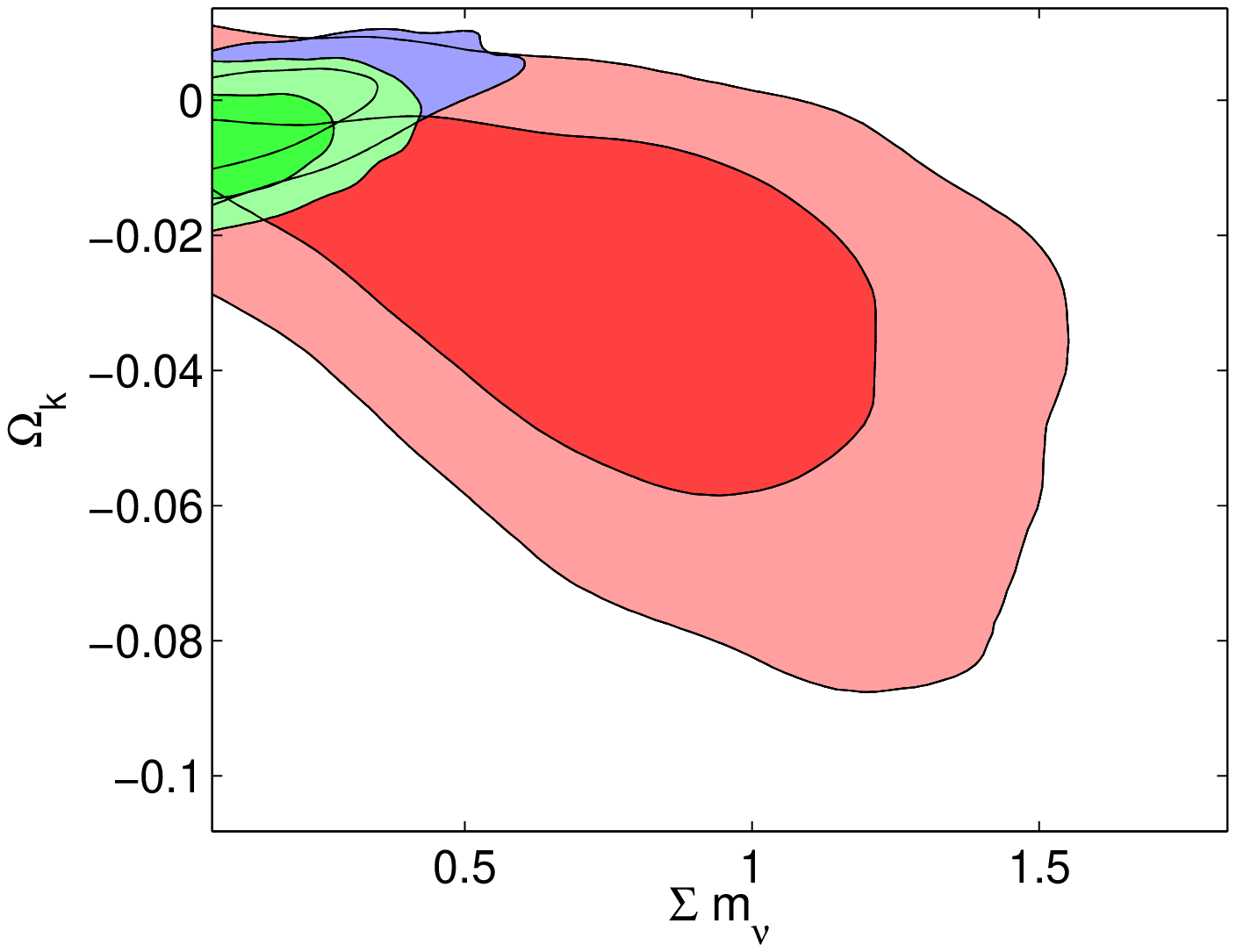}\\
\end{tabular}
 \caption{Left panel: the red contours show the $68\%$ and $95\%$~CL allowed
  regions from the PLANCK data set in the ($\sum m_\nu$, $w$)
  plane, while the blue and green contours show the impact of
  the addition of the DR9 BAO signature and the full shape of DR9 galaxy
  clustering measurements respectively. The magenta contours depict the combination
  of PLANCK with DR9 galaxy clustering data and SNLS3 measurements.
  Right panel: as in the left panel but
  in the ($\sum m_\nu$, $\Omega_k$) plane (note the absence 
  of the case with SNLS3 data in the analyses presented in this figure).}
\label{fig:w}
\end{figure*}
\subsection{Dark energy and massive neutrinos}
In this section we explore the bounds on the sum of neutrino masses if the
dark energy equation of state $w$ is allowed to vary, ($w$CDM model). There exists a strong and very well
known degeneracy in the $\sum m_\nu - w$ plane~\cite{hannestad}.
If the neutrino mass is allowed to freely vary, the amount of cold dark
matter is required to increase in order to leave the
matter power spectrum unchanged. This change of  $\Omega_m$  can also occurr
if $w$ is allowed to freely vary as well. Consequently, cosmological
neutrino mass bounds will become weaker if the dark
energy equation of state is included as a free parameter. 
Table~\ref{tab:constraints2} presents the galaxy clustering limits on
the sum of neutrino masses and on the dark energy equation of state
$w$ within the $w$CDM scenario. For the sake
of comparison, we depict as well the constraints from the PLANCK data set
alone. The addition of HST data to the basic PLANCK CMB data set barely changes
the $95\%$~CL constraint of $\sum m_\nu<1$~eV. While the addition of
DR8 BOSS data, neither in the form of clustering measurements, nor in the
form of geometrical BAO constraints changes these limits significantly~\footnote{Without shot noise parameters the
addition of DR8 angular power spectrum to Planck data results in a
much better constraint than the one quoted in
Tab.~\ref{tab:constraints2}, being $\sum m_\nu <0.77$~eV at
$95\%$~CL.}, the addition of the DR9 3D power spectrum
measurements sets a $95\%$~CL limit of $\sum m_\nu< 0.48$~eV. 
This limit is much better than the one provided by the combination of DR9
BAO information~\cite{anderson} and the PLANCK data set in a $w$CDM
universe, which is $\sum m_\nu < 0.71$~eV at $95\%$~CL.

Concerning $w$, the mean values and  the $95\%$~CL associated  errors
depicted in Tab.~\ref{tab:constraints2} show that the combination of
galaxy clustering measurements with the 
PLANCK CMB data set is not able to extract $w$ with high precision: the constraints we
obtained from this data combination for $w$ are rather weak but perfectly consistent with a
$\Lambda$CDM model. The addition of Supernovae Ia luminosity distance measurements
from the 3 year Supernova Legacy Survey (SNLS3)~\cite{snls3} reduces significantly
the errors on the dark energy equation of state: the combination
of PLANCK plus SNLS3 provides a mean value and $95\%$~CL errors on the
dark energy equation of state parameter of $w=-1.21^{+0.20}_{-0.22}$. If DR9
galaxy clustering data is also added in the analysis, $w=-1.16^{+0.15}_{-0.17}$.

Figure~\ref{fig:w}, left panel, shows the $68\%$ and $95\%$~CL allowed
regions in the ($\sum m_\nu$, $w$) plane from the PLANCK data set
described in Sec.~\ref{sec:data}, and also from the combination of the
former data set  with DR9 BAO geometrical information and with DR9 galaxy clustering (i.e. full shape)
measurements. Notice that the neutrino mass limits using the galaxy clustering
information are better than those obtained using the BAO
signature alone. Indeed, DR9 BAO measurements show a mild preference for
$w<-1$, allowing therefore for a larger neutrino mass. 
 We also investigate the impact of adding Supernovae
Ia luminosity distance constraints to the combination of PLANCK and DR9 galaxy
clustering data sets: while the impact on the sum of the neutrino mass
bound is negligible, the errors on the dark energy equation of state
parameter $w$ are reduced by a factor of three. 

\begin{table*}
\begin{center}
\begin{tabular}{lcccc}
\hline \hline
             &      Planck+WP +lensing   &       Planck+WP +lensing
             &   Planck+WP +lensing\\
               &  (+HST)          &       +DR8 (+HST)  &  +DR9 (+HST)\\
\hline
\hspace{1mm}\\
${\mnu}[eV]$  & $<1.01$  $(0.97)$ & $<1.02$  $(0.95)$ & $<0.48$  $(0.58)$ \\
\\
$w$        &  $-1.55_{-0.45}^{+0.54}$    & $-1.42_{-0.58}^{+0.49}$ & $-1.10_{-0.57}^{+0.44}$\\
\\
              &  $(-1.53_{-0.45}^{+0.37})$   &    $(-1.54_{-0.37}^{-0.45})$  &$(-1.30_{-0.34}^{+0.30})$\\
\\
\hline
\hline
\end{tabular}
\caption{$95\%$~CL upper bounds on ${\mnu}$ from the different data combinations
  considered here within a $w$CDM model, with (without) the HST prior on the Hubble constant
$H_0$. We show as well the mean value of $w$ together with its
$95\%$~CL errors. The results with DR8 (DR9) data sets refer
to the case in which the full-shape of the angular (3D) power spectrum
is considered, including shot noise parameters (systematic
corrections) in the analyses. The constraint from the full shape of
DR9 galaxy clustering measurements is highly superior to that arising from the combination of DR9
BAO information~\cite{anderson} and the PLANCK data set in a $w$CDM
universe, which is $\sum m_\nu < 0.71$~eV at $95\%$~CL.}
\label{tab:constraints2}
\end{center}
\end{table*}

\subsection{Curvature and massive neutrinos}
We present here the constraints on neutrino masses in the context of a
non flat universe, allowing for a non negligible curvature component,
see Tab.~\ref{tab:priors} for the priors adopted in the curvature
component. 
Table~\ref{tab:constraints3} shows our constraints for the  PLANCK 
data set, PLANCK plus DR8 angular power spectrum data and PLANCK plus DR9 galaxy clustering
measurements with and without a prior on the Hubble constant
$H_0$ from HST. In this non flat model, DR8 angular clustering
measurements combined with PLANCK reduce the constraint on
$\sum m_\nu$, from $\sum m_\nu<1.36$~eV to $\sum m_\nu<0.92$~eV (both at $95\%$~CL). This constraint
is very similar to the one obtained if the BAO DR8 geometrical
information is used, $\sum m_\nu <0.80$~eV.
Adding the HST prior to DR8 angular power spectrum measurements
improves significantly  the constraints: the $95\%$~CL upper limit is $\sum m_\nu< 0.33$~eV.

DR9 3D power spectrum measurements greatly improve 
the results from the PLANCK  data set: when combined with our basic PLANCK
dataset, the $95\%$~CL bounds without the HST prior are $\sum m_\nu<
0.35$~eV with systematic uncertainties. If HST data is included as well in the analysis, the former $95\%$~CL
bound translates into $\sum m_\nu< 0.26$~eV. These limits are better than those
obtained from the combination of the PLANCK data set with the DR9 BAO
measurement, which is $\sum m_\nu< 0.47$~eV without the HST prior. 
Therefore, this non flat model,  together with the $w$CDM one, is a working
example in which constraints from full shape 3D power-spectrum
measurements provide significant extra information than those from BAO
signature alone.

Figure~\ref{fig:w}, right panel, shows the $68\%$ and $95\%$~CL allowed
regions in the ($\sum m_\nu$, $\Omega_k$) plane from the PLANCK data
set described in Sec.~\ref{sec:data}, and from the combination of the former data set with DR9 BAO
measurements, and DR9 galaxy clustering information. Notice that the neutrino mass constraint arising from
the clustering measurements is more powerful than those obtained
exploiting the BAO signature.

Concerning $\Omega_k$, the mean value and the associated $95\%$~CL
errors are not significantly changed when galaxy clustering 
measurements are included.

\begin{table*}
\begin{center}
\begin{tabular}{lcccc}
\hline \hline
      &             Planck+WP+lensing     &     Planck+WP+lensing
      &   Planck+WP+lensing\\
      &             (+HST)          &    +DR8 (+HST)  &  +DR9 (+HST)\\
\hline
\hspace{1mm}\\
${\mnu}[eV]$  & $<1.36$  $(0.32)$ & $<0.92$  $(0.33)$ & $<0.35$  $(0.26)$ \\
\\
$\Omega_{\rm k}$       &  $-0.031_{-0.041}^{+0.036}$  & $-0.01_{-0.019}^{+0.018}$ & $0.005_{-0.009}^{+0.01}$\\
\\
  &   $0.007_{-0.010}^{+0.009}$  & $(0.006_{-0.010}^{+0.010})$ & $(0.001_{-0.009}^{+0.008})$\\
\\
\hline
\hline
\end{tabular}
\caption{$95\%$~CL upper bounds on ${\mnu}$ in a non-flat model from the different data combinations
  considered here, with (without) the HST prior on the Hubble constant
$H_0$. We depict as well the mean value and the $95\%$~CL errors for the curvature
energy density $\Omega_k$. The results with DR8 (DR9) data sets refer
to the case in which the full-shape of the angular (3D) power spectrum
is considered, including shot noise parameters (systematic
corrections) in the analyses. The neutrino mass bound extracted from
the full shape measurements of BOSS DR9 are better than the one
obtained using the DR9 BAO measurement~\cite{anderson}, which is $\sum m_\nu< 0.47$~eV
at $95\%$~CL without the HST prior.}
\label{tab:constraints3}
\end{center}
\end{table*}

\section{Conclusions}
\label{sec:concl}
Cosmology provides an independent laboratory to test physical
properties of fundamental particles. Neutrino masses affect the different cosmological
observables in different ways, and therefore it is possible to
derive strong constraints on the sum of their masses by combining
different cosmological data sets. Cosmic Microwave Background physics is affected by the
presence of massive neutrinos via the \emph{Early Integrated Sachs
Wolfe effect}, since the transition from the relativistic to the non
relativistic neutrino regime will induce a non trivial evolution of
the metric perturbations. Massive neutrinos will also suppress the
lensing potential.

Large scale structure measurements of the
galaxy power spectrum are affected by massive neutrinos, since
they are hot relics with large velocity dispersion which, at a given
redshift, erase the growth of matter perturbations on spatial scales smaller than the
typical neutrino free streaming scale.
Recent measurements of the Planck CMB experiment do not provide a strong bound 
on the sum of the neutrino masses. The addition of a prior on the Hubble constant from
the Hubble Space Telescope improves the results in a very
significant way since it breaks the strong degeneracy between the
neutrino mass and the Hubble constant. 
However, Planck and HST data sets show some tension in the measurement
of the Hubble parameter. While Baryon Acoustic Oscillation
measurements also improve the neutrino mass bounds
 when combined with Planck data, it is crucial to explore if
measurements using the full shape of the matter power spectrum can
further improve the neutrino mass limits, in particular, in non
minimal cosmological scenarios with a curvature or with a dark energy
equation of state $w\neq -1$. 

Here we combine recent Planck data with galaxy power spectrum measurements
from the BOSS experiment, one of the four surveys of
the Sloan Digital Sky Survey III,
(SDSS-III) to derive the constraints on the sum of neutrino masses. We explore both the 2D angular power spectrum measurements
from the CMASS sample of luminous
galaxies of SDSS-III Data Release 8 (DR8) as well the full 3D power spectrum
shape of SDSS-III Data Release 9 (DR9).

In the context of a minimal $\Lambda$CDM scenario, DR9 3D galaxy
clustering measurements
 improve significantly the existing neutrino mass bounds from Planck data.
We find $\sum m_\nu< 0.39$~eV at $95\%$ confidence level for the
combination of the DR9 3D
power spectrum with Planck CMB data (with lensing included) and
Wilkinson Microwave Anisoptropy Probe 9-year polarization
measurements. Similar results are obtained with the DR9 BAO geometrical signature.
Therefore, the $95\%$ confidence level constraint of $\sum m_\nu<1.1$~eV
obtained in the absence of large scale structure measurements is
greatly improved and robust neutrino mass constraints can be obtained
without the addition of the controversial prior on the Hubble constant from HST.  

In the presence of a freely varying (constant) dark energy equation of
state, we find $\sum m_\nu<0.49$~eV at $95\%$ confidence level for the
combination of the DR9 3D
power spectrum with Planck CMB data (with lensing included) and
Wilkinson Microwave Anisoptropy Probe 9-year polarization
measurements, making this constraint highly superior to that obtained
when replacing galaxy clustering data by the HST prior. 

In non flat geometries, the combination of the DR9 3D
power spectrum with Planck CMB data (with lensing included) and
Wilkinson Microwave Anisoptropy Probe 9-year polarization
measurements provides the neutrino mass bound $\sum m_\nu<0.35$~eV at $95\%$ confidence level.
If we use instead the associated DR9 BAO geometrical info, the $95\%$
confidence level neutrino mass bounds in the $w$CDM and non flat cosmologies 
are $\sum m_\nu<0.71$~eV and $\sum
m_\nu<0.46$~eV, respectively.  Consequently, in extended
cosmological scenarios with a free dark energy equation of state 
or with a curvature component, measurements of the full shape of the
galaxy power spectram are extremely helpful, providing better results
than those obtained with the associated Baryon Acoustic Oscillation
signature only.

While we were completing this study, a new
analysis~\cite{Riemer-Sorensen:2013jsa} combining Planck data, galaxy
clustering measurements from the WiggleZ Dark Energy Survey and other
external data sets has appeared in the literature. The bound is $\sum
m_\nu<0.24$~eV at $95\%$ confidence level when combining Planck with
WiggleZ power spectrum measurements, setting $k_{\rm{max}}=0.2$~$h$/Mpc.  
The analyses presented here are however penalized by our large
systematic uncertainties: as a comparison,
when we neglect in our analyses systematic uncertainties, we get $\sum
m_\nu<0.25$~eV at $95\%$ confidence level after combining Planck with
DR9 galaxy clustering measurements (with a shot noise nuisance
parameter included and setting $k_{\rm{max}}=0.12$~$h$/Mpc).

\section{Acknowledgments}
The authors would like to thank Shun Saito for help concerning DR9
power spectrum measurements and Maria Archidiacono for useful help
with the manuscript. Part of the research described in this paper was carried out at the Jet
Propulsion Laboratory, California Institute of Technology, under a contract
with the National Aeronautics and Space Administration. O.M. is
supported by the Consolider Ingenio 
project CSD2007-00060, by PROMETEO/2009/116, by the Spanish Ministry Science project FPA2011-29678 and by the ITN Invisibles PITN-GA-2011-289442.




\end{document}